\documentclass[12pt,psfig]{article}
\usepackage{amssymb}
\usepackage{amsmath,amsthm}
\usepackage{amsfonts}
\usepackage{graphicx}
\usepackage{braket}
\usepackage{graphics}
\usepackage{indentfirst}
\numberwithin{equation}{section}

\usepackage{hyperref}

\begin{document}
\begin{center}\Large\textbf{Radiation of Closed Strings
between the Parallel Dynamical-Dressed Unstable
D$p$-Branes}
\end{center}
\vspace{0.75cm}
\begin{center}
{\large Hamidreza Daniali and \large Davoud Kamani}
\end{center}
\begin{center}
\textsl{\small{Department of Physics, Amirkabir University of
Technology (Tehran Polytechnic) \\
P.O.Box: 15875-4413, Tehran, Iran \\
e-mails: hrdl@aut.ac.ir , kamani@aut.ac.ir \\}}
\end{center}
\vspace{0.5cm}

\begin{abstract}

We introduce a boundary state which is corresponding
to a D$p$-brane with tangential dynamics in the presence
of the Kalb-Ramond field, a tachyonic field and a $U(1)$
gauge potential in a special gauge. From the interaction
of such branes radiation amplitude of a general massless
closed string will be computed. The effects of the large
distances of the branes on this radiation will be studied.
In the large distances of the branes, the possibility
of axion radiation will be investigated. Our calculations
will be in the context of the bosonic string theory.

\end{abstract}

{\it PACS numbers}: 11.25.-w; 11.25.Uv

\textsl{Keywords}: Boundary state; Background fields;
Dynamics; Radiation amplitude; Axion production.

\newpage
\section{Introduction}

By studying the D-branes properties and
their interactions, some of the
important physical results
in the string theory were obtained \cite{1,2}.
One of the convenient methods for calculating the
interaction of two branes, in the closed-channel, is
the boundary state formalism \cite{3}-\cite{15}.
The interaction of two bare-static D-branes
in the superstring theory
is zero \cite{16}.
By adding dynamics and various background
fields, some nonzero and valuable interaction amplitudes
were obtained \cite{17}-\cite{27}.

Closed strings can be radiated from
the D$p$-branes with various
configurations. One of the most important setups
is the radiation of closed strings from a single
unstable D$p$-brane \cite{28, 29, 30}.
Besides the tachyonic background, which
prominently induces instability
on the branes, closed string radiation
from a single unstable
brane in the presence of the various background fields
has been also investigated \cite{31, 32}.
In addition, the supersymmetric version of the
closed string radiation was also constructed
\cite{33}. Another interesting
configuration is the closed string radiation
from the interacting D-branes. This kind of radiation was
studied only in specific setups \cite{34, 35}.
We shall add dynamics, internal and background fields
to the interacting unstable branes to
extract the closed string radiation from a generalized
configuration.

The background fields and dynamics of the branes
motivated us to obtain the effects of these variables on
the closed string radiation from the interacting branes.
Therefore, in this paper, in the context of the bosonic
string theory, we investigate the massless closed string
radiation from the interaction of two unstable
D$p$-branes with tangential dynamics in the
presence of the Kalb-Ramond field, a
quadratic tachyonic field and
a U(1) gauge potential in a special gauge.
For computing the radiation amplitude, we shall
apply the boundary state formalism.
Thus, by inserting an appropriate
vertex operator in the worldsheet of the exchanged
closed string between the branes,
we produce a radiated closed string.
We accurately use the eikonal approximation in
which the recoil of the branes is ignored.
After calculations for radiating a general
massless state, we concentrate on the axion radiation
from the distant branes. We shall observe that
the radiated axion is directly emitted
by one of the two interacting branes.
Note that the interaction between two D-branes in the
large distance limit occurs only via the exchange of
the massless states.

This paper is organized as follows. In Sec. \ref{200},
the boundary state, corresponding to a dynamical-dressed
unstable D$p$-brane, will be introduced.
In Sec. \ref{301}, the radiation amplitude of a general
massless closed string from the interaction of two
parallel D$p$-branes will be constructed.
In Sec. \ref{303}, the foregoing radiation amplitude
will be deformed for the distant branes.
In Sec. \ref{400}, the radiation of an axion
from the distant branes will be obtained.
Sec. \ref{500} is devoted to the conclusions.

\section{The boundary state of the
dynamical-dressed unstable D$p$-brane}
\label{200}

In this section we introduce the boundary state
corresponding to a dynamical-dressed unstable D$p$-brane.
Hence, we start with the following sigma-model
action for closed string
\begin{eqnarray}
\label{2.1}
S =&-&\frac{1}{4\pi\alpha'} {\int}_\Sigma
{\rm d}^{2}\sigma \left(\sqrt{-g}g^{ab}
G_{\mu\nu}\partial_a X^{\mu}\partial_b
X^{\nu}+\varepsilon^{ab} B_{\mu\nu}
\partial_a X^{\mu}\partial_b
X^{\nu}\right)
\nonumber\\
&+&\frac{1}{2\pi\alpha'} {\int}_{\partial\Sigma}
{\rm d}\sigma \left( A_\alpha
\partial_{\sigma}X^{\alpha}+
\omega_{\alpha\beta}J^{\alpha\beta}_{\tau }
+T(X^\alpha)\right),
\end{eqnarray}
where $\mu,\nu \in \{0,1,\cdots , 25\} $ represent
the spacetime indices,
which are split into the indices for
the worldvolume directions, i.e.
$\alpha, \beta \in \{0 , \cdots ,p \}$,
and for the perpendicular directions to it, i.e.
$i, j \in \{p+1 , \cdots ,25 \}$.
Besides, $a,b\in\{0,1\}$
specify the worldsheet directions.
$\Sigma$ is the closed string worldsheet,
while $\partial\Sigma$ is its boundary.
$G_{\mu\nu} $ and $g_{ab}$ are the metrics of the
target spacetime and string worldsheet, respectively.
The background fields are the Kalb-Ramond field
$B_{\mu\nu}$, the $U(1)$ internal gauge field $A_\alpha$
and the open string tachyon field $T(X^\alpha)$.
The constant antisymmetric tensor $\omega_{\alpha \beta}$
shows the angular velocity of the brane. Hence,
the $\omega$-term obviously expresses the tangential
dynamics of the brane, whose explicit form is
given by ${\omega }_{\alpha \beta}J^{\alpha\beta }_{\tau }=
2{\omega }_{\alpha \beta}X^{\alpha}{\partial }_{\tau
}X^{\beta}$. Besides the tangential dynamics, one can
also impose a transverse dynamics to the brane.
This can be solely exerted by
the boost transformations on the boundary state equations
\cite{22}, \cite{32}, \cite{34},
\cite{36}, \cite{37}, \cite{38}.
However, for simplicity, we shall consider only
the tangential motion and rotation.

We apply the flat spacetime with the metric
$ G_{\mu\nu} = \eta_{\mu\nu}={\rm diag}(-1,+1,\cdots,+1)$
and a constant Kalb-Ramond field $B_{\mu\nu}$. Besides,
we utilize the trusty gauge
$A_{\alpha}=-\frac{1}{2}F_{\alpha \beta }X^{\beta}$
with the constant field strength
$F_{\alpha \beta }$. In addition, we use the
quadratic tachyon
profile $T= \frac{1}{2}U_{\alpha\beta}X^{\alpha}X^{\beta}$
where $U_{\alpha\beta}$ is a constant symmetric matrix.
We should note that the origin and the
conformal invariance of the action \eqref{2.1}
have been widely studied in various papers, e.g., see
the Refs. \cite{3}, \cite{39}, \cite{40},
\cite{41}, \cite{42}, \cite{43}
and \cite{44} (and also references therein).

By setting the variation of the action to zero
we receive the equation of motion and
the following boundary state equations
\begin{eqnarray}
\label{2.2}
&~& \left[{\mathbf A}_{\alpha \beta}
{\partial}_{\tau }X^{\beta}
+{\mathcal F}_{\alpha \beta}
{\partial }_{\sigma }X^{\beta}
+U_{\alpha \beta }X^{\beta }\right]_{\tau =0}\ \ |B\rangle\ =0,
\nonumber\\
&~& {\delta X}^i|_{\tau =0}|B\rangle\ =0,
\end{eqnarray}
where the total field strength is
${\cal{F}}_{\alpha \beta}=F_{\alpha \beta}-B_{\alpha \beta}$
and ${\mathbf A}_{\alpha \beta}=\eta_{\alpha \beta}
+ 4\omega_{\alpha \beta}$.
For the next purpose we bring in the solution of the
equation of motion of $X^\mu(\sigma,\tau)$,
\begin{eqnarray}
\label{2.3}
X^\mu (\sigma ,\tau )=x^{\mu}+2 \alpha^\prime Q^{\mu }\tau
+\frac{i}{2} \sqrt{2 \alpha^\prime}
\sum_{m\ne 0}\frac{1}{m}\left( {\alpha }^{\mu }_m
e^{-2im\left(\tau -\sigma \right)}
+{\widetilde{\alpha }}^{\mu }_m e^{-2im(\tau +\sigma )}
\right).
\end{eqnarray}

In fact, because of the presence of the
background fields on the brane worldvolume,
the Lorentz symmetry has been clearly broken. Thus,
the tangential dynamics of the
brane along its worldvolume directions is sensible.
Now we prove this.
The effect of the Lorentz generators on the
boundary state can be obtained from Eq. \eqref{2.2},
\begin{eqnarray}
J^{\alpha\beta} |B\rangle &=& \int_0^\pi
d\sigma \bigg[(\mathbf A^{-1}
\mathcal{F})^\alpha_{\ \gamma} X^\beta
\partial_\sigma X^\gamma
- (\mathbf A^{-1} \mathcal{F})^\beta_{\ \gamma}
X^\alpha \partial_\sigma X^\gamma
\nonumber\\
&+& (\mathbf A^{-1} U)^\alpha_{\ \gamma}
X^\beta X^\gamma
- (\mathbf A^{-1} U)^\beta_{\ \gamma}
X^\alpha X^\gamma\bigg]_{\tau=0} |B\rangle .
\label{2.04}
\end{eqnarray}
This equation prominently demonstrates that
for restoring the Lorentz
symmetry, the tachyon matrix $U_{\alpha \beta}$
and the total field strength $\mathcal{F}_{\alpha\beta}$
should vanish. We see that even in the absence of the
electric and magnetic fields, the tachyon field independently
breaks the Lorentz symmetry along the worldvolume of the brane.
However, since the RHS of Eq. \eqref{2.04} depends on the
spacetime coordinates along the brane worldvolume we conclude
that the Lorentz symmetry breaking is local.
Therefore, the internal linear
motion and rotation of the brane in any direction
clearly are sensible.

For receiving more perception of the tangential dynamics
of a D$p$-brane, beside the Lorentz symmetry
breaking, we should note that such setups
are T-dual of some imaginable systems.
In other words, a
D$p$-brane with the tangential dynamics
can be constructed via T-duality
from a D$(p-1)$-brane which
moves and rotates parallel and perpendicular
to its volume. For example, consider a D-string
along the direction $x^1$ with the velocity
components $V^1$ and $V^2$ in the $x^1$-
and $x^2$-directions, respectively.
Now apply the T-duality in
the direction $x^2$, in which we suppose
that it is compact.
The resultant system is a D2-brane, which has been expanded
along the plane $x^1x^2$ with the velocity $V^1$ in the
direction $x^1$ and an electric field $E=V^2$
in the direction $x^2$. In the same way, one can
acquire a rotating D2-brane through the T-duality from another
setup of a D-brane. Note that after exerting
the T-duality all compact coordinates can be decompactified.

Combining Eqs. \eqref{2.2} and \eqref{2.3},
the boundary state equations are conveniently expressed
in terms of the zero modes and oscillators.
The coherent state method enables us to
obtain the solution of the oscillatory
part of the boundary state equations
\begin{eqnarray}
\label{2.4}
{|B\rangle}^{({\rm osc})}\ =\prod^{\infty }_{n=1}
{[\det Q_{(n)}]^{-1}}{\exp \left[-\sum^{\infty }_{m=1}
{\frac{1}{m}\left({\alpha }^{\mu }_{-m}S_{(m)\mu \nu }
{\widetilde{\alpha }}^{\nu }_{-m}\right)}\right]\ }
{|0\rangle}_{\alpha }
\otimes{|0\rangle}_{\widetilde{\alpha }}\;,
\end{eqnarray}
where the normalization factor
$\prod^{\infty }_{n=1}{{[\det Q_{(n)}]}^{-1}}$
comes from the disk partition function.
In fact, the advent of this prefactor is
due to the square structure of the
action \eqref{2.1}. That is, in
addition to the gauge choice
$A_{\alpha}=-\frac{1}{2}F_{\alpha \beta }X^{\beta}$,
the other terms in the action also possess the
squared form of $X^\alpha$s.
Thus, this configuration path-integrally is feasible
and the normalizing prefactor must be inserted \cite{11}.
The matrices possess the following definitions
\begin{eqnarray}
&~& Q_{(m){\alpha \beta }} =
{\mathbf A}_{\alpha \beta}-
{{\mathcal F}}_{{\mathbf \alpha }{\mathbf \beta
}}+\frac{i}{2m}U_{\alpha \beta },\\
&~& S_{(m)\mu\nu}=(\Delta_{(m)\alpha \beta}\;
,\; -{\delta}_{ij}),\label{2.6}\\
&~& \Delta_{(m)\alpha \beta} =
(Q_{(m)}^{-1}N_{(m)})_{\alpha \beta},\\
&~& N_{(m){\alpha \beta }} = {\mathbf A}_{\alpha \beta}
+{{\mathcal F}}_{{\mathbf \alpha }{\mathbf \beta }}
-\frac{i}{2m}U_{\alpha \beta },
\end{eqnarray}
For the zero-mode part of the boundary state
we receive the following expression
\begin{eqnarray}
\label{2.10}
{{\rm |}B\rangle}^{\left(0\right)}&=&
\prod_i{\delta {\rm (}x^i}{\rm -}y^i{\rm )}
{\rm |}p^i{\rm =0}\rangle\;
\int^{\infty }_{{\rm -}\infty }
{\prod^p_{\alpha=0}\bigg{\{}{\rm d}p^{\alpha }}
\exp\Bigg[i{\alpha }^{{\rm '}}\Bigg(
{\left(U^{{\rm -}{\rm 1}}
{\mathbf A}\right)}_{\alpha \alpha }
{\left(p^{\alpha }\right)}^{{\rm 2}}
\nonumber\\
&+&\sum^{}_{\beta \ne \alpha}
{{\left(U^{{\rm -}{\rm  1}}
{\mathbf A}+{\mathbf A}^T U^{-1}\right)}_{\alpha \beta }
p^{\alpha }p^{\beta }}\Bigg)\Bigg]{\rm \ \ }
{\rm |}p^{\alpha }\rangle
\bigg{\}}. \ \ \ \
\end{eqnarray}

By considering the contribution of the conformal
ghosts to the boundary state, which is
\begin{eqnarray}
{|B\rangle}^{(\rm gh)}
=\exp\left[\sum^{\infty }_{m=1}
{\left(c_{-m}{\tilde{b\ }}_{-m}-b_{-m}
{\tilde{c}}_{-m}\right)}\right]\frac{c_0+{\tilde{c}}_0}{2}
|q=1\rangle\ \otimes|\tilde{q}=1\rangle,
\end{eqnarray}
one can construct the total boundary state
\begin{eqnarray}
{|B\rangle}^{(\rm tot)}=\frac{T_p}{2}
{|B\rangle}^{(\rm osc)}\otimes
{|B\rangle}^{(0)}\otimes{|B\rangle}^{(\rm gh)},
\end{eqnarray}
where $T_p$ is the D$p$-brane tension.

\section{Radiation of a general massless closed string}
\label{300}

In this section we calculate the
radiation amplitude of a general massless closed string
from the interaction of two parallel
dynamical-dressed unstable D$p$-branes.
To generalize our calculations
let us assume that the fields and dynamics of the two
interacting branes to be different. Therefore,
the subscripts (1) and (2) will be used
to exhibit these differences.

\subsection{The radiation amplitude}
\label{301}

In the closed string channel, the
interaction of two D-branes
takes place by exchanging a closed
string between the branes.
The geometry of the exchanged string
worldsheet is a cylinder with $\tau$ as the
coordinate along the length of the
cylinder, $ 0\le \tau \le t$, and $\sigma$ as the periodic
coordinate,  i.e. $ 0 \le \sigma\le \pi$.
The radiation of a closed string state
is elaborated by inserting the corresponding vertex operator
of the string state, i.e. $V(\tau, \sigma)$,
into the amplitude of the interaction.
Precisely, one should compute
\begin{equation}
\label{3.1}
\mathcal{A} = \int_{0}^{\infty} dt \int_{0}^{t}
d\tau \ ^{(\rm tot)}\langle B_1|
e^{-tH_{\rm (tot)}} V(\tau, \sigma) |B_2\rangle^{(\rm tot)},
\end{equation}
where $H_{\rm (tot)}$ comprises the ghost and matter
parts of the closed string Hamiltonian
\begin{eqnarray}
H_{\rm (tot)}=H_{\rm ghost}+{\alpha'}
Q^{{\rm2}}{\rm +\ 2}\sum^{\infty
}_{n{\rm =1}}{{\rm (}{\alpha }_{{\rm -}n}
\cdot {\alpha }_n}{\rm +}
{\widetilde{\alpha }}_{{\rm -}n}\cdot
{\widetilde{\alpha }}_n{\rm)-} 4.
\end{eqnarray}
Let us apply $z=\sigma +i \tau$ and $\partial = \partial_z$.
Subsequently, the vertex operator for a general
massless string possesses the feature
\begin{equation}
V(z, \bar{z}) = \epsilon_{\mu\nu} \partial X^\mu
\bar{\partial} X^\nu e^{ip\cdot X},
\end{equation}
where $\epsilon_{\mu\nu}$ is the
polarization tensor and $p^\mu$ (with
$p^\mu p_\mu =0 $) is the momentum
of the radiated closed string.

Since each string coordinate is the summation of
the zero-mode part and oscillating portion, i.e.
$ X^\mu (\sigma, \tau) = X^\mu_{0}(\tau)
+ X^\mu_{{\rm osc}} (\sigma, \tau)$, we can write
\begin{eqnarray}\label{3.4}
{\partial} X^\mu \bar{\partial} X^\nu e^{ip\cdot X}
&=& \big( {\partial} X^\mu_{0} \bar{\partial} X^\nu_{0}
+{\partial} X^\mu_{0} \bar{\partial}X^\nu_{{\rm osc}}
\nonumber\\
&+& {\partial} X^\mu_{{\rm osc}} \bar{\partial} X^\nu_{0} +
{\partial} X^\mu_{{\rm osc}}
\bar{\partial} X^\nu_{{\rm osc}}\big)
e^{ip\cdot X_{0}} e^{ip\cdot X_{{\rm osc}}}.
\end{eqnarray}
This implies that the appearing
terms in the vertex operator can be written in
the general form
\begin{equation}
\label{3.5}
\epsilon_{\mu\nu}\left[\Gamma(X_{0})
\Lambda ({X_{{\rm osc}}})\right]^{\mu\nu}
e^{ip\cdot X_{0}} e^{ip\cdot X_{{\rm osc}}},
\end{equation}
where $\Gamma(X_{0}) \in
\{1,\; \partial X_{0},\;\bar{\partial}X_{0},\;
\partial X_{0}\bar{\partial}X_{0}\}$
and $\Lambda({X_{{\rm osc}}}) \in
\{1,\; \partial X_{\rm osc},
\;\bar{\partial}X_{\rm osc},\;
\partial X_{\rm osc}\bar{\partial}X_{\rm osc}\}$.
This notation allows us to
separate the computation as follows
\begin{eqnarray}
\label{3.6}
^{(\rm tot)}\langle B_1| e^{-tH_{\rm tot}} V(\tau, \sigma)
|B_2\rangle^{(\rm tot)} &=& \frac{T_p^2}{4}
\epsilon_{\mu\nu} \Big[  \ ^{(0)}\langle B_1 | e^{-tH(X_{0})}
\Gamma(X_{0})e^{ip\cdot X_{0}} | B_2 \rangle^{(0)}
\nonumber\\
&\times&  ^{\rm{(osc)}}\langle B_1| e^{-t H_{\rm(osc)} }
\Lambda ({X_{\rm osc}}) e^{ip\cdot X_{\rm osc}}
|B_2\rangle^{\rm{(osc)}} \Big]^{\mu\nu}
\nonumber\\
&\times&  ^{\rm{(g)}}\langle B_1| e^{-t H_{\rm(g)} }
|B_2\rangle^{\rm{(g)}},
\end{eqnarray}
where $H(X_0) = \alpha' Q^2 -4$.

After very long calculations, for the various forms of
$\Gamma(X_{0})$, we receive
\begin{eqnarray}
^{(0)}\langle B_1 | e^{-tH(X_{0})} e^{ip\cdot X_{0}}
| B_2 \rangle^{(0)}
&=& (2\pi)^{26} \mathcal{D}(y_1,y_2) e^{4t}
\int_{-\infty}^{+ \infty}\int_{-\infty}^{+ \infty}
\prod_{\alpha=0}^{p} dk^\alpha dk^{\prime \alpha}
\nonumber \\
&\times&\mathfrak{D}(k; k^\prime)
e^{- \alpha'(l'k^2+\tau k^{\prime^2})} ,
\nonumber\\
^{(0)}\langle B_1 | e^{-tH(X_{0})} \partial X^\mu_{0}
e^{ip\cdot X_{0}} | B_2 \rangle^{(0)} &=& -
\alpha^\prime (2\pi)^{26} \mathcal{D}(y_1,y_2) e^{4t}
\int_{-\infty}^{+ \infty}\int_{-\infty}^{+ \infty}
\prod_{\alpha=0}^{p} dk^\alpha dk^{\prime \alpha}
\nonumber \\
&\times& \mathfrak{D}(k; k^\prime)
k^\mu  e^{- \alpha'(l'k^2+\tau k^{\prime^2})} ,
\nonumber\\
^{(0)}\langle B_1 | e^{-tH(X_{0})}
\bar{\partial} X^\mu_{0}
e^{ip\cdot X_{0}} | B_2 \rangle^{(0)} &=&
\alpha^\prime (2\pi)^{26} \mathcal{D}(y_1,y_2) e^{4t}
\int_{-\infty}^{+ \infty}\int_{-\infty}^{+ \infty}
\prod_{\alpha=0}^{p} dk^\alpha dk^{\prime \alpha}
\nonumber \\ &\times&\mathfrak{D}(k; k^\prime)
 k^\mu e^{- \alpha'(l'k^2+\tau k^{\prime^2})} ,
\nonumber\\
^{(0)}\langle B_1 | e^{-tH(X_{0})} \partial X^\mu_{0}
\bar{\partial} X^\nu_{0} e^{ip\cdot X_{0}} | B_2 \rangle^{(0)}
&=& - \alpha^{\prime 2} (2\pi)^{26} \mathcal{D}(y_1,y_2) e^{4t}
\int_{-\infty}^{+ \infty}\int_{-\infty}^{+ \infty}
\prod_{\alpha=0}^{p} dk^\alpha dk^{\prime \alpha}
\nonumber \\
&\times& \mathfrak{D}(k; k^\prime)
k^\mu k^\nu
e^{- \alpha'(l'k^2+\tau k^{\prime^2})},
\end{eqnarray}
where, by applying the Wick's rotation
$\tau \rightarrow -i\tau$,
we introduced another proper
time $l' = t-\tau$. Additionally,
the functions $\mathcal{D}(y_1,y_2)$ and
$\mathfrak{D}(k; k^\prime)$ have the following definitions
\begin{eqnarray}
\mathcal{D}(y_1, y_2) &\equiv&
\prod_{i=p+1}^{25} \delta(x^i - y_1^i)
\delta(x^i - y_2^i) \delta(p^i),
\label{3.8}\\
\mathfrak{D} (k;k') &\equiv&
\prod_{\alpha=0}^{p} \delta
\left( p^\alpha + k^{\prime\alpha} - k^\alpha\right)
\cr
&\times& \exp\Bigg[ -i \alpha^\prime
\Bigg( \sum_{\alpha = 0}^{p} \left[(U_{1}^{-1}
\mathbf{A}_1)_{\alpha\alpha} (k^\alpha)^2
-(U_{2}^{-1}\mathbf{A}_2)_{\alpha\alpha} (k'^\alpha)^2
\right]
\cr
&+& 2\sum_{\alpha\ne\beta}\left[(U_1^{-1} \mathbf{A}_1
)_{\alpha\beta}k^\alpha k^\beta
- (U_2^{-1} \mathbf{A}_2 )_{\alpha\beta}
k'^\alpha k'^\beta \right]
\Bigg) \Bigg] .
\quad\
\label{3.9}
\end{eqnarray}

In Eq. \eqref{3.6}, for the factor which includes
the oscillators, we write
\begin{eqnarray}
\label{3.10}
^{\rm(osc)}\langle B_1| e^{-t H{\rm(osc)} }
\Lambda ({X_{\rm osc}})
e^{ip\cdot X_{\rm osc}} |B_2\rangle^{\rm (osc)}
\equiv \langle \Lambda ({X_{\rm osc}})
e^{ip\cdot X_{\rm osc}}\rangle
Z^{\rm(osc)},
\end{eqnarray}
where $Z^{\rm (osc)}$ is the oscillation part of the
partition function
\begin{equation}
Z^{\rm (osc)} = \ ^{\rm(osc)}\langle B_1|
e^{-t H_{\rm(osc)} } |B_2\rangle^{\rm(osc)}.
\end{equation}
The other factor in Eq. \eqref{3.10} is given by
\begin{equation}
\label{3.12}
\langle \Lambda ({X_{\rm osc}})
e^{ip\cdot X_{\rm osc}}\rangle
\equiv \dfrac{^{\rm(osc)}\langle B_1|
e^{-t H_{\rm (osc)} }\Lambda ({X_{\rm osc}})
e^{ip\cdot X_{\rm osc}}
|B_2\rangle^{\rm(osc)}}{^{\rm(osc)}
\langle B_1| e^{-t H_{\rm (osc)} }
|B_2\rangle^{\rm(osc)}}\;.
\end{equation}
For the various forms of $\Lambda ({X_{\rm osc}})$
Eq. \eqref{3.12} finds the following features
\begin{eqnarray}
\langle \partial X^\mu_{\rm osc}
e^{ip\cdot X_{\rm osc}}\rangle &=&
i \langle \partial X^\mu p\cdot X \rangle_{\rm osc}
\langle e^{ip\cdot X_{\rm osc}} \rangle,\label{3.13}\\
\langle \bar{\partial} X^\mu_{\rm osc}
e^{ip\cdot X_{\rm osc}}\rangle
&=& i \langle \bar{\partial} X^\mu p\cdot X
\rangle_{\rm osc} \langle
e^{ip\cdot X_{\rm osc}} \rangle,\label{3.14}\\
\langle \partial X^\mu_{\rm osc}
\bar{\partial} X^\nu_{\rm osc}
e^{ip\cdot X_{\rm osc}}\rangle
&=& \left[ \langle \partial X^\mu
\bar{\partial} X^\nu \rangle_{\rm osc}
- \langle \partial X^\mu p . X\rangle_{\rm osc} \langle
\bar{\partial} X^\nu p\cdot X \rangle_{\rm osc} \right]
\langle e^{ip\cdot X_{\rm osc}} \rangle.\ \ \
\end{eqnarray}

The new proper time $l' =t-\tau$ enables
us to change the integrations as
\begin{equation}
\int_{0}^{\infty}dt \int_{0}^{t} d\tau
= \int_{0}^{\infty} d\tau \int_{0}^{\infty} dl^\prime.
\end{equation}
In fact, $l^\prime$ and $\tau$ indicate the proper times
of the radiated closed string from the right (first) brane
and the left (second) brane, respectively. Therefore,
$\tau=0$ ($l^\prime =0$) specifies the
radiation from the first (the second) brane,
i.e. that brane with the boundary state $|B_1 \rangle$
($|B_2 \rangle$). The case
$\tau, l^\prime > 0$ obviously corresponds to
radiation that has been occurred between the
branes.

Adding all these together we can
write the general radiation amplitude as
\begin{eqnarray}
\label{3.17}
\mathcal{A} &=& \dfrac{T_p^2}{4} (2\pi)^{26}
\int_{0}^{\infty}d\tau \int_{0}^{\infty}  d l^\prime \
\mathcal{D}(y_1, y_2) e^{4(l' +\tau)}
\int_{-\infty}^{+ \infty}
\int_{-\infty}^{+ \infty}
\prod_{\alpha=0}^{p} dk^\alpha dk^{\prime \alpha}
\mathfrak{D} (k;k^\prime)
\nonumber \\
&\times& e^{- \alpha'(l'k^2+\tau k^{\prime^2})}
Z^{\rm (osc,g)} \langle
e^{ip\cdot X_{osc}} \rangle \mathcal{M}, \ \ \
\end{eqnarray}
where $\mathcal{M}$ and $Z^{\rm (osc,g)}$ are given by
\begin{eqnarray}
\label{3.18}
\mathcal{M} &=& \epsilon_{\mu\nu}
\Big\{\langle \partial X^\mu \bar{\partial} X^\nu
\rangle_{\rm osc} - \langle \partial
X^\mu p . X\rangle_{\rm osc}
\langle \bar{\partial} X^\nu p\cdot X \rangle_{\rm osc}
- i \alpha^\prime k^\mu
\langle \bar{\partial} X^\nu p\cdot X \rangle_{\rm osc}
\nonumber \\
&+& i \alpha^\prime k^\nu
\langle \partial X^\mu p\cdot X \rangle_{\rm osc} -
\alpha^{\prime 2}k^\mu k^\nu \Big\},
\end{eqnarray}
\begin{eqnarray}
Z^{\rm (osc,g)}= \ ^{\rm(osc)}\langle B_1|
e^{-(\tau + l') H_{\rm(osc)}}
|B_2\rangle^{\rm(osc)}\;\;
^{\rm{(g)}}\langle B_1|
e^{-(\tau + l')H_{\rm(g)} }
|B_2\rangle^{\rm{(g)}}.
\end{eqnarray}
According to the factor
$e^{- \alpha'(l'k^2+\tau k^{\prime^2})}$
the integration over $l'$ and $\tau$ leads to
the factors similar to $1/k^2$ or $1/k'^2$ or both, which
correspond to the propagators of the
radiated massless strings by the branes.

\subsection{The correlators}
\label{302}

Now we should compute all correlators in Eq. \eqref{3.17}.
In Eqs. (3.7) we employed the Wick's rotation
to determine the zero-mode correlators.
Now we should extract the oscillatory correlators
in this frame.
Thus, after some heavy calculations we acquire
\begin{eqnarray}
\label{3.20}
Z^{\rm (osc,g)} &=& \prod_{n=1}^{\infty}
\det \left[ Q^{\dagger}_{(n)  1} Q_{(n)  2}\right]^{-1} \;
\prod_{n=1}^{\infty} \dfrac{(1-q^{2n})^2}{\det
\left({\mathbf 1}- S^{(1)\dagger}_{(n)}
S^{(2)}_{(n)} q^{2n} \right) }\;,
\end{eqnarray}
\begin{eqnarray}
\langle \partial X^\mu X^\nu \rangle_{\rm osc}
&=& i \alpha^\prime \sum_{n=1}^{\infty}
\Bigg\{\eta^{\mu\nu}\text{Tr} \left(
\dfrac{{\mathbf 1}
+ S^{(1) \dagger}_{(n)}S^{(2)}_{(n)} q^{2n} }
{{\mathbf 1} - S^{(1) \dagger}_{(n)}
S^{(2)}_{(n)} q^{2n}}\right)
\nonumber \\
&-& S^{(1) \dagger\mu\eta}_{(n)} S^{(2)\nu}_{(n) \ \eta}
\text{Tr} \left( \dfrac{ S^{(1) \dagger}_{(n)}S^{(2)}_{(n)}
q^{2n} }{{\mathbf 1} - S^{(1) \dagger}_{(n)}
S^{(2)}_{(n)} q^{2n}}\right)
\nonumber\\
&-& S^{(2) \mu\nu}_{(n)}
\text{Tr} \Bigg[ \dfrac{S^{(1) \dagger}_{(n)}
S^{(2)}_{(n)} q^{2(n-1)} e^{-4l^\prime}}{{\mathbf 1}
- S^{(1) \dagger}_{(n)}
S^{(2)}_{(n)} q^{2(n-1)} e^{-4l^\prime}}
\nonumber\\
&\times& \left( {\mathbf 1} + \dfrac{1}{{\mathbf 1}
- S^{(1) \dagger}_{(n)}S^{(2)}_{(n)}
q^{2(n-1)} e^{-4l^\prime}}\right)\Bigg]
\nonumber\\
&+& S^{(1) \dagger \mu\nu}_{(n)}
\text{Tr} \Bigg[ \dfrac{S^{(1) \dagger}_{(n)}
S^{(2)}_{(n)}
q^{2(n-1)} e^{-4\tau}}{{\mathbf 1}- S^{(1) \dagger}_{(n)}
S^{(2)}_{(n)} q^{2(n-1)} e^{-4\tau}}
\nonumber\\
&\times&\left( {\mathbf 1}+ \dfrac{1}{{\mathbf 1}- S^{(1)
\dagger}_{(n)}S^{(2)}_{(n)} q^{2(n-1)}
e^{-4\tau}}\right)\Bigg]
\Bigg\},\label{3.21} \ \ \ \ \ \ \ \
\end{eqnarray}
\begin{eqnarray}
\langle \partial X^\mu \bar{\partial} X^\nu \rangle_{\rm osc}
&=& 2 \alpha^\prime \sum_{n=1}^{\infty}
\Bigg\{ 2 n \eta^{\mu\nu}\text{Tr}
\left( \dfrac{S^{(1) \dagger}_{(n)}S^{(2)}_{(n)}
q^{2n}}{\left({\mathbf 1}- S^{(1) \dagger}_{(n)}
S^{(2)}_{(n)} q^{2n}\right)^2}\right)
\nonumber\\
&-& n S^{(1) \dagger\mu\eta}_{(n)}
S^{(2)\nu}_{(n) \ \eta}
\text{Tr} \left( \dfrac{ S^{(1) \dagger}_{(n)}
S^{(2)}_{(n)}q^{2n} }
{\left({\mathbf 1} - S^{(1) \dagger}_{(n)}
S^{(2)}_{(n)} q^{2n}\right)^2}\right)
\nonumber\\
&+& 2 n S^{(2) \mu\nu}_{(n)}
\text{Tr} \left( \dfrac{S^{(1) \dagger}_{(n)}
S^{(2)}_{(n)}q^{2(n-1)} e^{-4l^\prime}}{
\left({\mathbf 1}- S^{(1) \dagger}_{(n)}S^{(2)}_{(n)}
q^{2(n-1)} e^{-4l^\prime}\right)^3}\right)
\nonumber\\
\qquad\qquad&+& 2 n
S^{(1)\dagger \mu\nu}_{(n)} \text{Tr}
\left( \dfrac{S^{(1) \dagger}_{(n)}S^{(2)}_{(n)}
q^{2(n-1)} e^{-4\tau}}{\left({\mathbf 1}
- S^{(1) \dagger}_{(n)}
S^{(2)}_{(n)}q^{2(n-1)}
e^{-4\tau}\right)^3}\right)\Bigg\},
\label{3.22}
\end{eqnarray}
where $ q=e^{-2(\tau + l')}$. One can conveniently show
that $ \langle \partial X^\mu X^\nu \rangle_{\rm osc}
= - \langle \bar{\partial} X^\mu X^\nu \rangle_{\rm osc}$.

Using the Cumulant expansion, the exponential
$\langle e^{ip\cdot X_{\rm osc}}\rangle$
can be elaborated in terms of the correlators of $X$'s.
According to the boundary states formulation
and Eq. \eqref{3.12}, the correlators
with odd number of $X$'s
vanish. Hence, only the correlators with
even number of $X$'s remain.
Now let us assume that the momentum of
the radiated string is small.
This implies that in the Cumulant expansion
one should only compute the factor
$\exp\left(-\frac{1}{2} p_\mu p_\nu \langle
X^\mu X^\nu\rangle_{\rm osc}\right)$.
Subsequently, for the massless radiated
strings, the following result is received
\begin{eqnarray}
\langle e^{ip.X_{\rm osc}} \rangle
&=& \prod_{n=1}^{\infty} \bigg{\{}
\det \left( \mathbf{1} - S^{(1) \dagger}_{(n)}
S^{(2)}_{(n)}\; q^{2n}
\right)^{ \frac{\alpha^\prime}{2n} p_\mu p_\nu
S^{(1) \dagger\mu\eta}_{(n)}S^{(2)\nu}_{(n) \ \eta}}
\nonumber\\
&\times&
\det \left(\mathbf{1} - S^{(1) \dagger}_{(n)}S^{(2)}_{(n)}
\;q^{2(n-1)} e^{-4 l^\prime}
\right)^{-\frac{\alpha^\prime}{2n}p_\mu
p_\nu S^{(2) \mu\nu}_{(n)}}
\nonumber\\
&\times&
\det \left(\mathbf{1} - S^{(1) \dagger}_{(n)}S^{(2)}_{(n)}
\;q^{2(n-1)}
e^{-4 \tau}\right)^{- \frac{\alpha^\prime}{2n}p_\mu p_\nu
S^{(1)\dagger \mu\nu}_{(n)}}
\nonumber\\
&\times& \det \left[ \exp
\left( \frac{\alpha^\prime p_\mu p_\nu S^{(2) \mu\nu}_{(n)}}
{2n}  \left(\mathbf{1} - S^{(1) \dagger}_{(n)}S^{(2)}_{(n)}
\;q^{2(n-1)} e^{-4 l^\prime}\right)^{-1} \right) \right]  \ \ \
\nonumber\\
&\times& \det \left[ \exp
\left( \frac{\alpha^\prime p_\mu p_\nu
S^{(1)\dagger \mu\nu}_{(n)}}{2n}
\left(\mathbf{1} - S^{(1) \dagger}_{(n)}S^{(2)}_{(n)}
\;q^{2(n-1)} e^{-4 \tau}\right)^{-1} \right) \right]
\bigg{\}}. \ \ \ \ \
\qquad\qquad\label{3.23}
\end{eqnarray}

\subsection{The large distance branes}
\label{303}

Here we consider radiation of the closed strings from the
branes which are located far from each other.
In this case only the exchange of the
massless closed strings possesses the dominant
contribution to the interaction.
Since the long enough time corresponds to the
large distance of the branes,
the limit $t \to \infty$ should merely exert on
the oscillating part of the amplitude, i.e. on
Eqs. \eqref{3.20}-\eqref{3.23}. Accordingly,
$Z^{\rm (osc,g)}$ and
the correlators in the amplitude \eqref{3.17}
find the following forms
\begin{eqnarray}
Z^{\rm (osc,g)}|_{t\rightarrow\infty}
= \prod_{n=1}^{\infty}
\det \left[ Q^{\dagger}_{(n)1} Q_{(n)2}\right]^{-1},
\label{3.24}
\end{eqnarray}
\begin{eqnarray}
\langle \partial X^\mu X^\nu
\rangle_{\rm osc}|_{t\rightarrow\infty}
&=& i \alpha^\prime
\Bigg\{-13 \eta^{\mu\nu}
\nonumber\\
&-& S^{(2) \mu\nu}_{(1)}
\text{Tr} \Bigg[ \dfrac{S^{(1) \dagger}_{(1)}
S^{(2)}_{(1)} e^{-4l^\prime}}{{\mathbf 1}
- S^{(1) \dagger}_{(1)}
S^{(2)}_{(1)} e^{-4l^\prime}}
\left( {\mathbf 1} + \dfrac{1}{{\mathbf 1}
- S^{(1) \dagger}_{(1)}S^{(2)}_{(1)}
e^{-4l^\prime}}\right)\Bigg]
\nonumber\\
&+& S^{(1) \dagger \mu\nu}_{(1)}
\text{Tr} \Bigg[ \dfrac{S^{(1) \dagger}_{(1)}
S^{(2)}_{(1)}
e^{-4\tau}}{{\mathbf 1}- S^{(1) \dagger}_{(1)}
S^{(2)}_{(1)} e^{-4\tau}}\left( {\mathbf 1}
+ \dfrac{1}{{\mathbf 1}- S^{(1)
\dagger}_{(1)}S^{(2)}_{(1)}
e^{-4\tau}}\right)\Bigg]
\Bigg\},\label{3.25} \ \ \ \ \ \ \ \
\end{eqnarray}
\begin{eqnarray}
\langle \partial X^\mu \bar{\partial} X^\nu
\rangle_{\rm osc}|_{t\rightarrow\infty}
&=& 4 \alpha^\prime \Bigg\{S^{(2) \mu\nu}_{(1)}
\text{Tr} \left( \dfrac{S^{(1) \dagger}_{(1)}
S^{(2)}_{(1)}e^{-4l^\prime}}{
\left({\mathbf 1}- S^{(1) \dagger}_{(1)}S^{(2)}_{(1)}
e^{-4l^\prime}\right)^3}\right)
\nonumber\\
\qquad\qquad
&+& S^{(1)\dagger \mu\nu}_{(1)} \text{Tr}
\left( \dfrac{S^{(1) \dagger}_{(1)}S^{(2)}_{(1)}
e^{-4\tau}}{\left({\mathbf 1}
- S^{(1) \dagger}_{(1)}S^{(2)}_{(1)}
e^{-4\tau}\right)^3}\right)\Bigg\},\label{3.26}
\end{eqnarray}
\begin{eqnarray}
\langle e^{ip.X_{\rm osc}}
\rangle|_{t\rightarrow\infty}
&=& \det \left[\left(\mathbf{1}
- S^{(1) \dagger}_{(1)}S^{(2)}_{(1)}
e^{-4 l^\prime}
\right)^{-\frac{\alpha^\prime}{2}p_\mu
p_\nu S^{(2) \mu\nu}_{(1)}}\right]
\nonumber\\
&\times& \det \left[\left(\mathbf{1}
- S^{(1) \dagger}_{(1)}S^{(2)}_{(1)}
e^{-4 \tau}\right)^{- \frac{\alpha^\prime}{2}
p_\mu p_\nu S^{(1)\dagger \mu\nu}_{(1)}}\right]
\nonumber\\
&\times& \exp \left[ \frac{\alpha^\prime}{2} p_\mu
p_\nu S^{(2) \mu\nu}_{(1)}
{\rm Tr}\left(\mathbf{1}
- S^{(1) \dagger}_{(1)}S^{(2)}_{(1)}
e^{-4 l^\prime}\right)^{-1} \right]  \ \ \
\nonumber\\
&\times& \exp \left[ \frac{\alpha^\prime}{2} p_\mu p_\nu
S^{(1)\dagger \mu\nu}_{(1)}{\rm Tr}
\left(\mathbf{1} - S^{(1) \dagger}_{(1)}S^{(2)}_{(1)}
e^{-4 \tau}\right)^{-1} \right]. \ \ \ \ \
\qquad\qquad
\label{3.27}
\end{eqnarray}
In fact, for obtaining Eq. \eqref{3.27} we assumed that
the parameters of the setup and the momentum of the
radiated closed string satisfy the condition
\begin{eqnarray}
p_\mu p_\nu \sum^\infty_{n=2} \left(
S^{(1)*}_{(n)} + S^{(2)}_{(n)}\right)^{\mu\nu}=0.
\end{eqnarray}
This implies that for the given setup parameters,
two components of the momentum of the radiated
closed string are specified in terms of the
other components.

Substitute Eqs. \eqref{3.24}-\eqref{3.27}
into Eq. \eqref{3.17}, the amplitude for the closed string
radiation between the large distance branes is acquired.

\section{The axion radiation}
\label{400}

In this section we shall accomplish the amplitude
of the Kalb-Ramond (axion)
radiation from the interacting distant branes. The
axion polarization tensor satisfies
$\epsilon_{\mu\nu} = - \epsilon_{\nu\mu} $ and
$ p^\mu \epsilon_{\mu\nu}  =0$.
Thus, Eq. \eqref{3.18} becomes
\begin{eqnarray}
\mathcal{M} &=& \epsilon_{\mu\nu}
\Biggl\{4 \alpha^\prime \left( S^{(2) \mu\nu}_{(1)}
K_{l^\prime} + S^{(1) \dagger \mu\nu}_{(1)}
K_{\tau}\right) - \alpha^{\prime 2}
\bigg[ p_\lambda p_\delta
\bigg(S^{(2) \mu \lambda}_{(1)} S^{(2) \nu\delta}_{(1)}
R_{l^\prime}^2
\nonumber \\
&-&\left( S^{(2) \mu \lambda}_{(1)}S^{(1)
\dagger \nu \delta}_{(1)}
+ S^{(1)\dagger \mu \lambda}_{(1)}
S^{(2) \nu \delta}_{(1)} \right)  R_\tau
R_{l^\prime}  + S^{(1)\dagger \mu \lambda}_{(1)}
S^{(1)\dagger\nu \delta}_{(1)} R_\tau^2 \bigg)
\nonumber\\
\qquad&+& k^\mu p_\lambda \left( S^{(2) \nu \lambda}_{(1)}
R_{l^\prime}
- S^{(1)\dagger\nu \lambda}_{(1)} R_\tau\right)
+ k^\nu p_\lambda \left( S^{(2)  \mu \lambda}_{(1)}
R_{l^\prime} - S^{(1)\dagger\mu \lambda}_{(1)}
R_\tau\right) \bigg] \Biggr\},
\ \  \ \ \ \label{4.1}
\end{eqnarray}
where we have defined
\begin{eqnarray}
R_{(l^\prime , \tau )} &=&
\text{Tr} \left[ \dfrac{S^{(1) \dagger}_{(1)}
S^{(2)}_{(1)}
e^{-4(l^\prime , \tau )}}{\mathbf{1}- S^{(1) \dagger}_{(1)}
S^{(2)}_{(1)}
e^{-4(l^\prime , \tau )}}
\left( \mathbf{1} + \dfrac{1}{\mathbf{1}
- S^{(1) \dagger}_{(1)}
S^{(2)}_{(1)} e^{-4(l^\prime , \tau )}}\right)\right],
\nonumber\\
K_{(l^\prime , \tau )}
&=& \text{Tr} \left[\dfrac{S^{(1) \dagger}_{(1)}
S^{(2)}_{(1)}e^{-4(l^\prime , \tau )}}
{\left(\mathbf{1}- S^{(1) \dagger}_{(1)}S^{(2)}_{(1)}
e^{-4(l^\prime , \tau )}\right)^3}\right].
\label{4.2}
\end{eqnarray}

For comparing our results with the Ref. \cite{34}
we can express $\mathcal{M}$ only in terms of the functions
$R_{(l' , \tau)}$. For this purpose
we impose the following condition on the
setup parameters
$\Delta^{(1)\dagger}_{(1)}\Delta^{(2)}_{(1)} =\mathbf{1}$.
Afterward, for large distance limit
the two-derivative term in Eq. \eqref{3.18}
can then be written as
\begin{eqnarray}
\langle \partial X^\mu \bar{\partial}
X^\nu \rangle_{\rm osc}|_{t\rightarrow \infty}
= - \langle \partial X^\mu X^\nu
\rangle_{\rm osc}|_{t\rightarrow \infty}
\left[ \langle p\cdot \partial X p\cdot X
\rangle_{\rm osc}|_{t\rightarrow \infty} +
\dfrac{i \alpha^\prime}{2} (k^2 - k^{\prime 2})\right].
\label{4.3}
\end{eqnarray}
Exerting Eq. \eqref{3.25} into the one-derivative terms
of Eq. \eqref{4.3}, we receive Eq. \eqref{3.18}
only in terms of the functions $R_{(l^\prime, \tau)}|_
{\Delta^{(1)\dagger}_{(1)}\Delta^{(2)}_{(1)} =\mathbf{1}}$.
Besides, we have the following identity \cite{34},\cite{35}
\begin{eqnarray}
0&=&\int_{0}^{\infty} {\rm d}\tau \partial_\tau
\left[e^{- \tau (\alpha^\prime k^{\prime^2} - 4)}
\exp\left(-\dfrac{1}{2}\alpha^\prime p_\mu p_\nu
S^{(1)\dagger \mu\nu}_{(1)} \text{Tr}
\left[\ln (\mathbf{1}-e^{-4\tau})-(\mathbf{1}
-e^{-4\tau})^{-1}\right]
\right)\right]
\nonumber\\
&=&\int_{0}^{\infty} {\rm d}\tau \Bigg{[}
\left(\alpha^\prime k^{\prime 2}-4
+ 2\alpha^\prime p_\mu p_\nu S^{(1)\dagger \mu\nu}_{(1)}
{\tilde R}_{\tau}\right)
e^{- \tau (\alpha^\prime k^{\prime^2} - 4)}
\nonumber\\
&\times&\exp\left(-\dfrac{\alpha^\prime p_\mu p_\nu
S^{(1)\dagger \mu\nu}_{(1)}}{2}
\text{Tr}\left[\ln (\mathbf{1}- e^{-4\tau})
- (\mathbf{1}-e^{-4\tau})^{-1}\right]\right) \Bigg{]}.
\end{eqnarray}
Similar identity can be obtained
for the integration over $ l^\prime $ with
${\tilde R}_{l^\prime}$, where
\begin{eqnarray}
{\tilde R}_{\tau}
&=& - \dfrac{\alpha^\prime k^{\prime 2} - 4}
{2 \alpha^\prime p_\mu p_\nu
S^{(1)\dagger \mu\nu}_{(1)}}\;,
\nonumber \\ {\tilde R}_{l^\prime}
&=& - \dfrac{\alpha^\prime k^{ 2} -4}
{2 \alpha^\prime p_\mu p_\nu S^{(2) \mu\nu}_{(1)}}\;.
\label{4.5}
\end{eqnarray}
Now we acquired $\mathcal{M}$ in terms of the functions
${\tilde R}_\tau$ and ${\tilde R}_{l'}$.

For receiving the final form of the axion amplitude
we should use the following integral too
\begin{eqnarray}
\int_{0}^{\infty}d\tau \int_{0}^{\infty}
d l^\prime e^{-l^\prime (\alpha^\prime k^2 - 4)}
e^{- \tau (\alpha^\prime k^{\prime^2} - 4)}
\langle e^{ip\cdot X_{osc}} \rangle
|_{\Delta^{(1)\dagger}_{(1)} \Delta^{(2)}_{(1)}
=\mathbf{1}}\approx \dfrac{1}{(\alpha^{\prime } k^2-4)
(\alpha^\prime k^{\prime 2 }-4)},
\end{eqnarray}
where for the correlator of the exponential
we used the following form
\begin{eqnarray}
\langle e^{ip\cdot X_{osc}} \rangle
|_{\Delta^{(1)\dagger}_{(1)}
\Delta^{(2)}_{(1)} =\mathbf{1}} &=& \exp\left\{
-\alpha^\prime p_\mu p_\nu S^{(2)
\mu\nu}_{(1)} \text{Tr}
\left[ \ln (\mathbf{1}- e^{-4 l^\prime})
- (\mathbf{1}-e^{-4 l^\prime})^{-1} \right] \right\}
\nonumber \\
&\times&\exp\left\{  -\alpha^\prime p_\mu p_\nu S^{(1)
\dagger \mu\nu}_{(1)}
\text{Tr} \left[ \ln (\mathbf{1}- e^{-4 \tau})
- (\mathbf{1}-e^{-4 \tau})^{-1} \right] \right\}.
\end{eqnarray}

Let us apply the constant shifts on the momenta, i.e.
$\mathcal{K}^\alpha = k^\alpha + l^\alpha $ and
$ \mathcal{K}^{\prime\alpha} = k^{\prime\alpha} + l^\alpha $,
in which the internal vector
$ l^\alpha$ satisfies the following conditions
\begin{eqnarray}
k.l = k^\prime.l=0\;,\;\;\;\;
l^2= -\frac{4}{\alpha^\prime}.
\end{eqnarray}
Adding all these together  we obtain the ultimate feature
of the amplitude as in the following
\begin{eqnarray}
\mathcal{A} &=& \dfrac{1}{4} (2\pi)^{26}T_p^2
\mathcal{D}(y_1, y_2)
\prod_{n=1}^{\infty}\left[
\det\left( Q^{\dagger}_{(n)1} Q_{(n)2}\right)
\right]^{-1}
\nonumber\\
&\times& \int_{-\infty}^{+ \infty}
\int_{-\infty}^{+ \infty}
\prod_{\gamma=0}^{p} d\mathcal{K}^\gamma
d\mathcal{K}^{\prime \gamma}
\mathfrak{D} (\mathcal{K};\mathcal{K}^\prime)
\epsilon_{\alpha\beta}
\left(\dfrac{\Gamma^{\alpha\beta}}{\mathcal{K}^{2}}
- \dfrac{\Upsilon^{\alpha\beta}}
{\mathcal{K}^{\prime 2}}\right),
\label{4.9}
\end{eqnarray}
where $\Gamma^{\alpha\beta}$ and
$\Upsilon^{\alpha\beta}$ are given by
\begin{eqnarray}
\Gamma^{\alpha\beta} &=& \dfrac{1}{p_\mu p_\nu
S^{(1)\dagger \mu\nu}_{(1)}} \left[p_\gamma
\left( (\mathcal{K} - l)^\alpha
\Delta^{(1)\dagger \beta \gamma}_{(1)}
+ (\mathcal{K} - l)^\beta
\Delta^{(1)\dagger \alpha \gamma}_{(1)}\right)
- \frac{1}{2}
\left(\mathcal{K}^2 -\mathcal{K}^{\prime 2}\right)
\Delta^{(1)\dagger \alpha \beta}_{(1)} \right]
\nonumber \\
&-& \dfrac{\mathcal{K}^{\prime 2}}{\left(p_\mu p_\nu
S^{(1)\dagger \mu\nu}_{(1)}\right)^2}
\left[ p_\gamma p_\eta\left(
\Delta^{(1)\dagger \alpha \gamma}_{(1)}
\Delta^{(1)\dagger \beta \eta}_{(1)}
- \Delta^{(1)\dagger \alpha \beta}_{(1)}
\Delta^{(1) \dagger\gamma \eta}_{(1)}\right)
+p^\lambda p_\lambda\Delta^{(1)\dagger \alpha\beta}_{(1)}
\right],
\nonumber\\
\Upsilon^{\alpha\beta} &=&
\dfrac{1}{p_\mu p_\nu S^{(2) \mu\nu}_{(1)}}
\left[ p_\gamma \left( (\mathcal{K} - l)^\alpha \Delta^{(2)
\beta \gamma}_{(1)}
+ (\mathcal{K} - l)^\beta
\Delta^{(2) \alpha \gamma}_{(1)}\right)
- \frac{1}{2} \left(\mathcal{K}^2
-\mathcal{K}^{\prime 2}\right)
\Delta^{(2) \alpha \beta}_{(1)} \right]
\nonumber\\
&+& \dfrac{\mathcal{K}^2}{\left(p_\mu p_\nu
S^{(2) \mu\nu}_{(1)}
\right)^2}
\left[ p_\gamma p_\eta
\left( \Delta^{(2) \alpha \gamma}_{(1)}
\Delta^{(2) \beta \eta}_{(1)}
- \Delta^{(2) \alpha \beta}_{(1)}
\Delta^{(2) \gamma \eta}_{(1)}\right)
- p^\lambda p_\lambda
\Delta^{(2)\alpha\beta}_{(1)} \right].
\end{eqnarray}
The indices $\alpha, \beta, \gamma, \lambda, \eta$
belong to the set $\{0,1,\cdots,p\}$ while $\mu,\nu$
are the spacetime indices.
Since our calculations were in the lowest order
Eq. \eqref{4.9} accurately is in the tree-level diagram.

The amplitude \eqref{4.9} manifestly represents the axion
radiation from the interaction of two parallel
dynamical-dressed unstable D$p$-branes
in the large distance. As we see, according to the
upper index $(1)$ ($(2)$) the tensor
$\Gamma^{\alpha\beta}$ ($\Upsilon^{\alpha\beta}$)
depends on the parameters of the first (second)
brane. Therefore, the first and second terms
of \eqref{4.9} elaborate
the axion radiation by the first and the second brane,
respectively. According to the factors in
Eq. \eqref{4.9}, a massless closed string state
is emitted by one of the branes,
then it is absorbed by the other brane, afterward
it travels as an excited state on the brane for a while,
it finally decays by radiating an axion state.
This radiation is a bremsstrahlung-like
process. We should mention that our results completely
are in accordance with the Ref. \cite{34}.

Note that in Eq. \eqref{4.9} the term
$1/\mathcal{K}^2 \mathcal{K}^{\prime 2}$
is absent, which clarifies that
there is no axion radiation near
the middle points between the branes.
For the distant branes
the middle region is far from the both branes.

In fact, in some physical quantities
the squared form of the amplitude,
i.e. $|\mathcal A|^2$, is appeared.
Let us call the amplitude \eqref{3.1} as
$\mathcal A_{12}$. By exchanging the branes
we receive $\mathcal A_{21}$.
We observe that the squared versions
of them are not equal, i.e.,
$|\mathcal{A}_{12}|^2\ne|\mathcal{A}_{21}|^2$.
This asymmetry precisely is due to the presence of the
vertex operator of the radiated closed string.
In other words, in the absence of any closed string
radiation we acquire
$|\mathcal{A}_{12}|^2 = |\mathcal{A}_{21}|^2$.
In this case we have only the pure interaction
between the branes.
The amplitude of the pure interaction is given
by Eq. \eqref{3.1} without the vertex operator and
the $\tau$-integration. Note that even for the
pure interaction there is
$\mathcal{A}_{12}\ne \mathcal{A}_{21}$.
Therefore, the amplitude (for the pure interaction
and / or radiation) is exactly elaborated by
$\mathcal{A}_{12}$, but not with the arithmetic mean
$\left(\mathcal{A}_{12} +\mathcal{A}_{21}\right)/2$,
e.g. see the Refs. \cite{3}-\cite{27}, \cite{34}.

\section{Conclusions}
\label{500}

We introduced a boundary state which is
corresponding to a dynamical
D$p$-brane in the presence of
the antisymmetric tensor field,
an internal $U(1)$ gauge potential
with constant field strength
and a specific open string tachyon field.
For computing the amplitude of closed string radiation
between two parallel D$p$-branes,
we combined the associated vertex operator of the
radiated string and the boundary state formalism.
We acquired the radiation amplitude for a general
massless closed string. Besides the arbitrary distance,
the amplitude for large distances of the branes was also
calculated. Presence of the various parameters
drastically generalized
the amplitude. By varying the parameters the value of the
radiation amplitude can
be accurately adjusted to any desirable value.

From the radiation amplitude of a general massless
state we explicitly computed the axion production
between the large distance D$p$-branes.
We observed that this radiation occurs only
from one of the branes. Finally, we demonstrated that
the axion production cannot occur
near the middle region between the branes. This is due to
the fact that the middle points are far from the
both branes.



\begin{thebibliography}{99}

\bibitem{1}
J.~Polchinski, “String Theory”, (Cambridge University Press,
Cambridge, 1998), Volumes I and II; C.~V.~Johnson, “D-Branes”,
(Cambridge University Press, Cambridge, 2003).

\bibitem{2}
J.~Polchinski,
Phys. Rev. Lett. \textbf{75}, 4724-4727 (1995).

\bibitem{3}
C. G. Callan, I. R. Klebanov,
Nucl. Phys. \textbf{B 465} (1996) 473.

\bibitem{4}
M.B. Green and P. Wai, Nucl. Phys.
\textbf{B431} (1994) 131.

\bibitem{5}
C. Bachas, Phys. Lett. \textbf{B 374} (1996) 37.

\bibitem{6}
M. Li, Nucl. Phys. \textbf{B 460} (1996) 351.

\bibitem{7}
M.B. Green and M. Gutperle, Nucl.
Phys. \textbf{B476} (1996) 484.

\bibitem{8}
M. Frau, A. Liccardo and R. Musto, Nucl. Phys.
\textbf{B 602} (2001) 39.

\bibitem{9}
F. Hussain, R. Iengo and C. Nunez,
Nucl. Phys. \textbf{B 497} (1997) 205.

\bibitem{10}
P. Di Vecchia, M. Frau, I. Pesando,
S. Sciuto, A. Lerda and R. Russo, Nucl.
Phys. \textbf{B507} (1997) 259.

\bibitem{11}
C. G. Callan, C. Lovelace, C. R. Nappi, S. A. Yost,
Nucl. Phys. \textbf{B 288} (1987) 525;
Nucl. Phys. \textbf{B 308} (1988) 221.

\bibitem{12}
O. Bergman, M. Gaberdiel and
G. Lifschytz, Nucl. Phys. \textbf{B509} (1998) 194.

\bibitem{13}
S. Gukov, I. R. Klebanov, A. M. Polyakov,
Phys. Lett. \textbf{B 423} (1998) 64.

\bibitem{14}
M. Bertolini, P. Di Vecchia, M. Frau, A. Lerda and R.
Marotta, Nucl. Phys. \textbf{B 621}
(2002) 157.

\bibitem{15}
P. Di Vecchia, A. Liccardo, R. Marotta and
F. Pezzella, Int. J. Mod. Phys. \textbf{A 20}
(2005) 4699-4796.

\bibitem{16}
J.~Polchinski,
``\textit{TASI lectures on D-branes},'' arXiv:hep-th/9611050;
\\
J.~Polchinski, S.~Chaudhuri and C.~V.~Johnson,
``\textit{Notes on D-branes},''
arXiv:hep-th/9602052.

\bibitem{17}
C.~Bachas,
Phys. Lett. \textbf{B} \textbf{374}, 37-42 (1996).

\bibitem{18}
P.~Di Vecchia and A.~Liccardo,
``\textit{D-branes in string theory. I}'',
NATO Sci. Ser. C \textbf{556}, 1-60 (2000).

\bibitem{19}
M. Billo, D. Cangemi, P. Di Vecchia,
Phys. Lett. \textbf{B 400} (1997) 63.

\bibitem{20}
L.~Girardello, C.~Piccioni and M.~Porrati,
Mod. Phys. Lett. \textbf{A} \textbf{19}, 2059-2068 (2004).

\bibitem{21}
J.~X.~Lu,
Nucl. Phys. \textbf{B} \textbf{934}, 39-79 (2018).

\bibitem{22}
F.~Hussain, R.~Iengo, C.~A.~Nunez and C.~A.~Scrucca,
Phys. Lett. \textbf{B} \textbf{409}, 101-108 (1997).

\bibitem{23}
M. Frau, I. Pesando, S. Sciuto, A. Lerda and R. Russo, Phys.
Lett. {\bf B 400} (1997) 52.

\bibitem{24}
H. Arfaei and D. Kamani, Phys. Lett. \textbf{B 452}
(1999) 54-60, arXiv:hep-th/9909167;
D. Kamani, Phys. Lett. \textbf{B 487} (2000) 187-191,
arXiv:hep-th/0010019;
E. Maghsoodi and D. Kamani,
Nucl. Phys. \textbf{B 922} (2017) 280-292,
arXiv:1707.08383 [hep-th];
D. Kamani, Europhys. Lett. \textbf{57} (2002) 672-676,
arXiv:hep-th/0112153.

\bibitem{25}
D. Kamani, Mod. Phys. Lett. \textbf{A 17} (2002) 237-243,
arXiv:hep-th/0107184;
S. Teymourtashlou and D. Kamani,
Eur. Phys. J. C {\bf 81}, 761 (2021),
arXiv:2108.10164 [hep-th];
D. Kamani, Eur. Phys. J. C {\bf 26}, 285-291 (2002),
arXiv:hep-th/0008020;
F. Safarzadeh-Maleki and D. Kamani,
Phys. Rev. \textbf{D 90} 107902 (2014),
arXiv:1410.4948 [hep-th].

\bibitem{26}
H. Arfaei and D. Kamani,
Nucl. Phys. \textbf{B 561} (1999) 57-76,
arXiv:hep-th/9911146;
D. Kamani, Nucl. Phys. {\bf B 601} (2001) 149-168,
arXiv:hep-th/0104089.

\bibitem{27}
D. Kamani, Annals of Physics \textbf{354} (2015) 394-400,
arXiv:1501.02453 [hep-th];
H. Arfaei and D. Kamani,
Phys. Lett. \textbf{B 475} (2000) 39-45,
arXiv:hep-th/9909079;
D. Kamani, Mod. Phys. Lett. \textbf{A 15} (2000) 1655-1664,
arXiv:hep-th/9910043.

\bibitem{28}
N.~D.~Lambert, H.~Liu and J.~M.~Maldacena,
JHEP \textbf{03}, 014 (2007).

\bibitem{29}
B.~Chen, M.~Li and F.~L.~Lin,
JHEP \textbf{11}, 050 (2002).

\bibitem{30}
I.~R.~Klebanov, J.~M.~Maldacena and N.~Seiberg,
JHEP \textbf{07}, 045 (2003).

\bibitem{31}
K.~Nagami,
JHEP \textbf{01}, 005 (2004).

\bibitem{32}
S.~J.~Rey and S.~Sugimoto,
Phys. Rev. \textbf{D} \textbf{67}, 086008 (2003).

\bibitem{33}
J.~Shelton, JHEP \textbf{01}, 037 (2005).

\bibitem{34}
F.~Hussain, R.~Iengo, C.~A.~Nunez and C.~A.~Scrucca,
Nucl. Phys. \textbf{B} \textbf{517} (1998), 92-124;
AIP Conf. Proc. \textbf{415}, (1998) 421,
arXiv:hep-th/9711021.

\bibitem{35}
J.~D.~Blum,
Phys. Rev. \textbf{D} \textbf{68}, 086003 (2003).

\bibitem{36}
A. Sen, Int. J. Mod. Phys. \textbf{A 20}, 5513 (2005).

\bibitem{37}
P. Di Vecchia and A. Liccardo,
``{\it D branes in string theory, II}'',
YITP Proceedings Series No. 4 (Kyoto,
Japan, 1999).

\bibitem{38}
M. Saidy-Sarjoubi and D. Kamani,
Phys. Rev. \textbf{D 92} (2015) 046003,
arXiv:1508.02084 [hep-th].

\bibitem{39}
R.~G.~Leigh, Mod. Phys. Lett. \textbf{A}4 2767 (1989).

\bibitem{40}
E.~Witten, Nucl. Phys. B \textbf{460}, 335-350 (1996).

\bibitem{41}
D.~Kutasov, M.~Marino and G.~W.~Moore,
JHEP \textbf{10}, 045 (2000).

\bibitem{42}
A.~A.~Tseytlin,
J. Math. Phys. \textbf{42}, 2854-2871 (2001).

\bibitem{43}
F.~Safarzadeh-Maleki and D.~Kamani,
Phys. Rev. \textbf{D 89} 026006 (2014),
arXiv:1312.5489 [hep-th].

\bibitem{44}
G.~Arutyunov, A.~Pankiewicz and B.~Stefanski, Jr.,
JHEP \textbf{06}, 049 (2001).

\end{thebibliography}
\end{document}